\documentstyle[11pt,newpasp,twoside,epsf]{article}
\def\edcomment#1{\iffalse\marginpar{\raggedright\sl#1\/}\else\relax\fi}
\reversemarginpar

\begin{document}
\title{Deep Near-Infrared Universe Seen in the Subaru Deep Field}
         
\author{Tomonori Totani} 

\affil{Princeton University Observatory, Peyton Hall, NJ08544, USA, and
         Theory Division, National Astronomical Observatory, Mitaka, Tokyo
         181-8588, Japan}

\begin{abstract}
The Subaru Deep Field provides the currently deepest $K$-selected sample of
high-$z$ galaxies ($K' \sim 23.5$ at 5$\sigma$). The SDF counts, colors, and
size distributions in the near-infrared bands are carefully compared with
pure-luminosity-evolution (PLE) as well as CDM-based hierarchical merging
(HM) models. The very flat faint-end slope of the SDF $K$ count indicates
that the bulk (more than 90\%) of cosmic background radiation (CBR) in this
band is resolved, even if we take into account every known source of
incompleteness. The integrated flux from the counts is only about a third of
reported flux of the diffuse CBR in the same band, suggesting that a new
distinct source of this missing light may be required. We discovered
unusually red objects with colors of $(J-K) \ga$ 3--4, which are even redder
than the known population of EROs, and difficult to explain by passively
evolving elliptical galaxies. A plausible interpretation, which is the only
viable one among those we examined, is that these are dusty starbursts at
high-$z$ ($z \sim 3$), whose number density is comparable with that of
present-day ellipticals or spheroidal galaxies, as well as with that of
faint submillimeter sources. The photometric redshift
distribution obtained by $BVRIz'JK'$ photometries is also compared with the
data, and the HM model is found to predict too few high-$z$ objects at $K'
\la 22$ and $z \la 2$; the PLE model with reasonable amount of absorption
by dust looks more consistent with the data. This result is apparently in
contradiciton with some previous ones for shallower observations, and we
discuss the origin of this.  These results raise a question for the HM
models: how to form massive objects with starbursts at such high redshifts,
which presumably evolve into present-day elliptical galaxies or bulges?
\end{abstract}

\section{Introduction}

The Subaru Deep Survey is a systematic project of the 8.2m Subaru telescope
to study the deep extragalactic universe. The Subaru Deep Field was selected
near the north Galactic pole, avoiding large Galactic extinction and nearby
galaxy clusters, and the airmass of this field is smaller than the Hubble
Deep Field at Mauna Kea (Maihara et al. 2000). The wide field near-infrared
(NIR) camera CISCO took a very deep NIR $2'\times 2'$ iamge in $J$ and $K'$
bands, with 5$\sigma$ magnitude limits of 25.1 and 23.5. This is the deepest
image in the $K$ band taken so far, providing a unique $K$-selected sample of
galaxies which should be useful for study of faint, high-$z$ galaxies. The
field was also deeply followed-up by optical instruments of FOCAS and
Suprime-Cam. Here we review some interesting implications obtained by these
data set, focusing on NIR galaxy counts, colors, and photometric-redshift
distribution, compared with some theoretical models of galaxy formation and
evolution. Although omitted here, some interesting results for the clustering
of Lyman break galaxies and Lyman alpha emitters at $z \sim 4$ have been
obtained in the SDF and another project of the Subaru/XMM-Newton deep survey,
thanks to the very wide field of the Suprime-Cam.  See Ouchi et al. (2001,
2002) for these.

\section{NIR Galaxy Counts and Contribution to CBR}
Figure 1 shows $K$ band SDF galaxy counts, compared with those estimated by
other observations (Totani et al. 2001a).  Here, we plot counts multiplied by
flux, rather than count itself, to show the contribution to the cosmic
background radiation (CBR) per magnitude. Both the raw and corrected counts
assuming point sources are showing very flat faint-end slope, with rapidly
decreasing contribution to CBR beyond $K \ga 18$. Therefore the extrapolation
of the galaxy counts into fainter magnitudes does not significantly increase
EBL but converges to a finite EBL flux, and this means that the bulk of EBL
from galactic light has already been resolved into discrete galaxies.  These
results require that the diffuse EBL in NIR bands should not be different
from the count integrations, provided that the ordinary galactic light is the
dominant source of the EBL in these bands, as generally believed.  However, a
few recently reported detections of diffuse EBL in these bands suggest that
the diffuse CBR flux in $K$ bands is consistently higher than the count
integrations by a factor of $\sim 3$: $\nu I_\nu = 27.8 \pm 6.7 \ \rm nW \
m^{-2} sr^{-1}$ (Cambr\'esy et al. 2001), $29.3 \pm 5.4$ (Matsumoto 2000),
and $20.2 \pm 6.3$ (Wright 2001), which should be compared with the
integration of $K$ counts ($\sim 8 \ \rm nW \ m^{-2} sr^{-1}$).

If the discrepancy between the diffuse EBL and count integration is
real, it might suggest the existence of very diffuse component which
is different from normal galaxies.
Before deriving this extraordinary conclusion, however, all possible
systematic uncertainties in the above estimates must extensively be
checked. One of such systematics is the contribution to EBL by the galaxies
missed in deep galaxy surveys. Since galaxies are extended sources,
the detectability near the detection limit is not as simple as 
point sources. Furthermore, the well-known effect of the cosmological
dimming of surface brightness [$S \propto (1+z)^{-4}$] should make 
high-$z$ galaxies very difficult to detect, while such objects
may have a significant contribution to EBL. The photometry scheme could
also be a problem, because there is considerable uncertainty
in the estimate of the magnitude of faint galaxies because of
`growing' the photometry beyond the outer detection isophotes of galaxies.

Therefore we estimated the contribution to CBR by galaxies missed in SDF,
based on a realistic theoretical model of galaxy counts which includes all
known physical or observational selection effects and incompleteness, based
on the method presented in Yoshii (1993) and Totani \& Yoshii (2000).  First,
we construct a model of galaxy counts which best fits to the observed raw
(i.e., uncorrected) counts, taking into account all the above selection
effects.  Then we can calculate the true galaxy counts and EBL flux using the
same model without selection effects, and comparison between the true counts
and observed counts gives an estimate of contribution by missing
galaxies. The results are shown in Fig. 1, where the dashed line is the
best-fit model to raw SDF counts taking into account the selection effects in
theoretical calculation, while the solid line is the same model but without
the selection effects.  By this procedure, we found that the correction by
the incompleteness would increase the CBR from galaxies by at most 10\%,
which is too small to reconcile the count integrations and diffuse CBR
measurements (Totani et al. 2001a). 
Therefore we conclude that there must be a new source of CBR in
the NIR band, which must be very different and distinct populations from
known galaxies, unless some unkown systematics have affected the diffuse CBR
measurements significantly.

\begin{figure}
\centerline{\epsfxsize=9cm \epsfbox{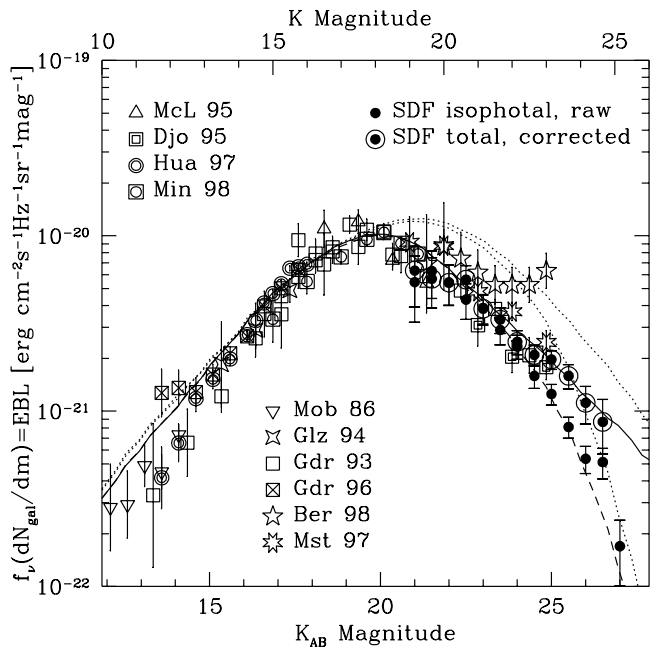}}
\caption{ The contribution to EBL by galaxies in the K band.
The filled circles are the raw SDF counts in
isophotal magnitude, while the symbols $\odot$ are the counts in total
magnitude which are corrected for incompleteness assuming point sources
(Maihara et al. 2000). The dashed line is the prediction by a PLE model for
which the selection effects under the observational conditions of SDF are
taken into account, fitting to the raw counts. The solid line is the same
prediction, but the selection effects are not included.  For detail, see
Totani et al. (2001a). The two dotted lines are model predictions with 
and without selection effects, using the same PLE model but including
a simple number evolution of $\eta =1$.}
\label{fig:ebl}
\end{figure}

The model shown in the dashed and solid lines is a so-called
pure-luminosity-evolution (PLE) model without number evolution, including
five types of galaxies (E/S0, Sab, Sbc, Scd, and Sdm), whose detail is given
in Totani \& Yoshii (2000). In fact, all the SDF NIR data of counts, $(J-K)$
colors, and size distributions are well described by this rather simple model
in the popular $\Lambda$-dominated flat universe, without any indication of
number evolution. This is somewhat in contrast to the result of the same
model against HDF galaxies, where a modest number evolution of $\eta \sim 1$
is required to fit the data [$\phi^* \propto (1+z)^\eta$, $L^* \propto
(1+z)^{-\eta}$] (Totani \& Yoshii 2000). This is the most naturally explained
by different merging histories for different galaxy types;
longer wavelengths are dominated by earlier types. Therefore,
these results indicate that elliptical galaxies which are dominant in the $K$
band are evolving without significant change of number density from $z \sim
2$ to the present (Totani et al.  2001c). A number evolution of $\eta \sim 1$
for elliptical galaxies is already inconsistent with the data, as shown by
dotted lines in Fig. 1.

On the other hand, a study of SDF K counts by a hierarchical merging (HM)
model based on CDM-based structure formation is presented in Nagashima 
et al. (2002), where the selection effects are carefuly taken into account
in a similar way. The HM also fits to the SDF counts in low density
cosmological models, but the PLE and HM models are giving different 
predictions in the redshift distribution, which will be discussed later.

\subsection{Unusually Red Objects}
An interesting discovery by SDF is the existence of unusually red objects in
NIR colors, with $J-K \ga $3--4. The brightest four objects of them are
presented in Maihara et al. (2000), and a more careful analysis has been
performed to estimate the number fraction of such objects as a function of
$K$ magnitude (Totani et al. 2001b). They found that the number fraction of
such objects sharply rises with increasing magnitude from $K \sim 20$,
reaching a few percent at the faintest magnitudes (see Fig. 2).  It should be
noted that such NIR color is even redder than the known population of the
extremely red objects (EROs), which are difined by red optical-NIR colors
such as $R - K > 5$; typical EROs have $J - K $ of at most 2. Furthermore,
such red NIR color cannot be explained by passively evolving elliptical
galaxies at any reasonable redshift without extinction by dust (Totani et
al. 2001b), while recent studies of EROs indicate that the majority of them
are passively evolving ellipticals (e.g., Daddi et al. 2000).

Then two interpretations remain for these hyper extremely red objects
(HEROs). One is ultra-high $z$ objects at $z \sim 10$, the red color being
due to the Lyman-break between $J$ and $K$ bands. Since HEROs are too faint to
measure redshifts, it is difficult to verify observationally.  However,
theoretically it is very unlikely; the rest-frame UV luminosity indicates
star formation rate of more than 100$M_\odot$/yr, but there should be very
few objects which are massive enough to allow such high SFR 
at $z \sim 10$, according to the widely believed CDM-based structure
formation theory. The other interpretation is that they are dusty starbursts,
which often show very red colors. In fact, Totani et al. (2001b) has shown
that the colors and counts of HEROs are well reproduced by a simple model if
present-day elliptical galaxies have formed by starbursts with a reasonable
amount of dust (i.e., inferred from model metallicity) at $z \sim 3$.

This redshift is similar to those estimated for the faintest submillimeter
sources in recent years. Interestingly, the counts of HEROs are roughly the
same with the faintest SCUBA sources, and submm flux expected by the
dusty-starburst model of HEROs is in fact close to the SCUBA sensitivity
limit. A detailed modeling of FIR-submm counts by Totani \& Takeuchi (2002)
has indeed shown that SCUBA counts are nicely explained by the dusty starbursts
of forming elliptical galaxies which are also responsible for HEROs.  These
results suggest an interesting possibility that HEROs and SCUBA sources are
the same population, which will evolve into present-day elliptical galaxies
or bulges.  It is very important to examine the correlation between these
two. Ultimate confirmation of this hypothesis would be brought in the NGST and
ALMA era.

\begin{figure}
\centerline{\epsfxsize=9cm \epsfbox{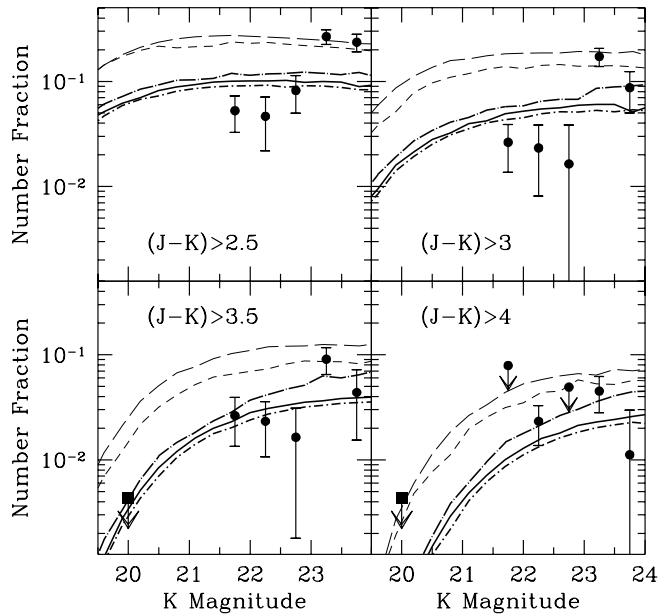}}
\label{fig:hero}
\caption{ 
Number fraction of galaxies redder than several threshold $J-K$ colors
(indicated in each panel), as a function of $K$ magnitude.  Filled circles are
the data of the SDF.  
The error bars are 1$\sigma$, while the upper limits shown by arrows
are at the 95\% confidence level. 
The upper limit at $K = 20$ is from Scodeggio \& Silva (2000, 
filled square).  The
solid line is the model prediction with the formation redshift $z_F = 3$ and
our standard dust-extinction normalization. See Totani et al. (2001b) for
the detail for other curves.
}
\end{figure}

\subsection{Photometric Redshift Distributions: PLE vs. HM}
Both the PLE and HM models fit to the SDF
$K$ galaxy counts. It may be because the galaxy counts do not have enough
power to discriminate these two, or may be because galaxies dominant in the
$K$ band in the HM model have only small or negligible number evolution. This
degeneracy can be broken by redshift distribution; especially a $K$-selected
sample has a strong power for the discrimination, since $K$-band light traces
the stellar mass which has formed so far, rather than star formation rate at
that time. Though spectroscopic redshifts are
unavailable for the faintest SDF galaxies, a reasonably reliable test is
possible by the photometric redshift technique, as is done by Kashikawa
et al. (2002). The result is shown in Fig. 3, in a form
of cumulative $z$-distribution separated by $K'$ magnitude intervals.

\begin{figure}
\centerline{\epsfxsize=9cm \epsfbox{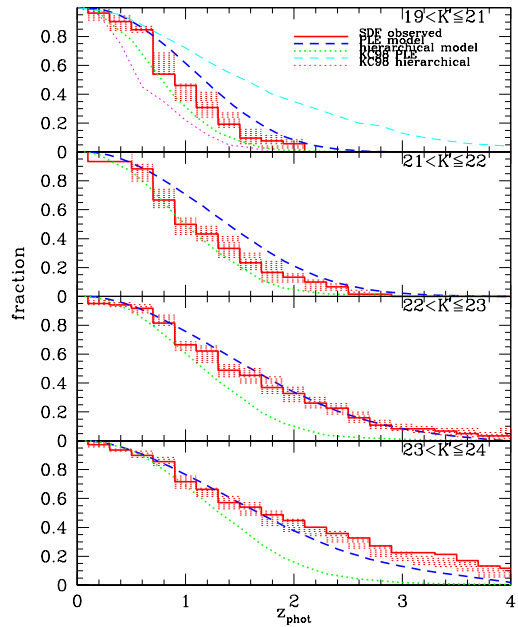}}
\caption{ The normalized cumulative redshift distribution of SDF sample
separated by $K'$ magnitude.  The thick solid histograms show the result of
the SDF photometric redshifts.  The shaded regions show $\pm 3\sigma$
deviated counts estimated by Monte Carlo realizations when photometric
redshift errors are taken into account. (Poisson fluctuation of number of
galaxies is not included.)  The thick dashed lines denote the prediction of
the PLE model in Totani et al. (2001), while the thick dotted lines for the
HM model in Nagashima et al. (2002). The predictions of dust-free PLE and HM
in Kauffman \& Charlot (1998) are also shown in the top panel by
thin dotted and dashed lines, respectively.  }
\end{figure}

In bright magnitude ranges of $K' < 22$, the observed distribution is
somewhat between the two models of PLE and HM. Although in the range of $21
\leq K' \leq 22$ the HM model looks better than the PLE, statistical
fluctuation is large because of the small number of galaxies.  The
Kolmogorov-Smirnov (KS) test gives a chance probability of 2 and 1\% of
getting the observed distribution from the PLE model. Considering model
uncertainties which is not taken into accoun in the KS test, it is impossible
to reject this model. On the other hand, the deficit of high-$z$ galaxies in
the HM model at $K' > 22$ is statistically much more significant (chance
probabilities of $6 \times 10^{-6}$ and $5 \times 10^{-15}$ at $22 \leq K'
\leq 23$ and $23 \leq K' \leq 24$, respectively). It should also be noted that
the HM model used here gives $z$-distribution more weighted to high redshifts
compared with another HM model of Kauffman \& Charlot (1998) (see the top
panel). We emphasize that this is a ``blind'' test, i.e., the models
shown here are those fitting best only to the SDF counts as described in
Totani et al. (2001c) and Nagashima et al. (2002) without further tuning of
parameters to the new photo-$z$ data of Kashikawa et al. (2002).

These results seem to be controversial when compared with several previous
papers doing a similar test (but in shallower magnitudes).  Fontana et
al. (1999) and Rudnick et al. (2001) claimed that the photometric redshift
distribution at $K < 21$ is consistent with HM models, but not with PLE
models. Firth et al. (2002) reached a similar conclusion using a sample of $H
< 20$. It should be noted that PLE models used by these groups are either
dust-free or assuming only constant extinction at the level of Galactic
extinction.  Such PLE models are inconsistent with the observed
$z$-distribution since they predict too many high-$z$ galaxies which are
visible because of strong starbursts assumed in the formation of elliptical
galaxies (see top panel of Fig. 3 for the dust-free PLE prediction by
Kauffmann \& Charlot 1998). However, dust-free model is obviously
unrealistic, especially for initial starbursts expected for elliptical
galaxies. We know that starbursting populations of galaxies quite often show
strong extinction and reddening. A chemical evolution model of elliptical
galaxies also suggests that the amount of metal produced in the initial
starburst phase is huge, and hence strong extinction seems quite plausible
(Totani \& Yoshii 2000). The PLE model used here (dashed line in Fig. 3) is
taking into account the extinction by a reasonable amount of dust inferred
from chemical evolution.  In addition, observational selection effects
discussed in \S 2 may also have affected previous results. (The theoretical
predictions by our PLE and HM models appropriately included all known
selection effects under the SDF condition).

Cimatti et al. (2002) compared a spectroscopic redshift distribution
at $K < 20$ obtained by the K20 survey with the latest PLE and HM models.
In fact, they found that, if dust extinction is taken into account
(another option is using the Scalo IMF rather than the Salpeter),
the difference between PLE and HM models becomes much smaller than
previously claimed. They found that such PLE models are in reasonable 
agreement with the data, but on the other hand, HM models predict too
many low redshift galaxies. 

\section{Concluding Remarks}
From these results, we conclude that the deepest $K$-selected sample of the
SDF is in overall agreement with the simple picture of PLE for early type or
elliptical galaxies, but the present version of HM models has a problem in
the redshift distribution of the faintest galaxies. Considering the overall
success of the CDM structure formation theory against various tests ${\it
not}$ based on galaxy luminosity and star formation activities (e.g.,
clustering properties or abundance of galaxy clusters), it is reasonable to
think that the problem identified for HM models is related with the treatment
of star formation activity.  It must incorporate a population of massive and
dusty starbursts at high redshift ($z \ga 3)$, which presumably evolved into
present-day elliptical galaxies or bulges without significant number
evolution.

The author would like to thank all the collaborators of the SDF project, and
the Subaru Telescope staffs who made this project possible. 
He has been financially supported in part by
the JSPS Fellowship for Research Abroad.

\end{document}